\documentclass[aps,prl,twocolumn]{revtex4-2}
\usepackage{amsmath}
\usepackage{amsfonts}
\usepackage{amssymb}
\usepackage{graphicx}
\usepackage[colorlinks,linkcolor=blue,anchorcolor=blue,citecolor=blue,urlcolor=blue]{hyperref}
\usepackage{mathrsfs}
\usepackage{epsfig}
\usepackage{float}
\usepackage{subfigure}
\usepackage{mathtools}
\usepackage{upgreek}
\usepackage{url}
\usepackage{gensymb}
\usepackage{colortbl}
\usepackage{xcolor}
\begin{document}
\bibliographystyle{apsrev4-2}

\title{Spectrally multiplexed and ultrabright entangled photon pairs in a lithium niobate microresonator}

\author{Bo-Yu Xu$^{1}$}
\thanks{These authors contributed equally to this work}
\author{Li-Kun Chen$^{2}$}
\thanks{These authors contributed equally to this work}
\author{Jintian Lin$^{3}$}
\thanks{These authors contributed equally to this work}
\author{Lan-Tian Feng$^{1,4}$}
\thanks{These authors contributed equally to this work}
\author{Rui Niu$^{1,4}$}
\author{Zhi-Yuan Zhou$^{1,4}$}
\author{Renhong Gao$^{3}$}
\author{Chun-Hua Dong$^{1,4}$}
\author{Guang-Can Guo$^{1,4}$}
\author{Qihuang Gong$^{2}$}
\author{Ya Cheng$^{3,5,6}$}
\email{ya.cheng@siom.ac.cn}
\author{Yun-Feng Xiao$^{2,6}$}
\email{yfxiao@pku.edu.cn}
\author{Xi-Feng Ren$^{1,4}$}
\email{renxf@ustc.edu.cn}

\affiliation{$^{1}$ CAS Key Laboratory of Quantum Information, University of Science and Technology of China, Hefei, Anhui 230026, P. R. China}
\affiliation{$^{2}$ State Key Laboratory for Mesoscopic Physics and Frontiers Science Center for Nano-optoelectronics, School of Physics, Peking University, Beijing 100871, China}
\affiliation{$^{3}$ State Key Laboratory of High Field Laser Physics and CAS Center for Excellence in Ultra-intense Laser Science, Shanghai Institute of Optics and Fine Mechanics (SIOM), Chinese Academy of Sciences (CAS), Shanghai 201800, China}
\affiliation{$^{4}$ CAS Synergetic Innovation Center of Quantum Information $\&$ Quantum Physics University of Science and Technology of China, Hefei, Anhui 230026, P. R. China}
\affiliation{$^{5}$ State Key Laboratory of Precision Spectroscopy, East China Normal University, Shanghai 200062, China}
\affiliation{$^{6}$ Collaborative Innovation Center of Extreme Optics, Shanxi University, Taiyuan 030006, China}

\maketitle

\noindent
\textbf{Abstract}\\
\noindent\textbf{
On-chip bright quantum sources with multiplexing ability are extremely high in demand for the integrated quantum networks with unprecedented scalability and complexity. 
Here, we demonstrate an ultrabright and broadband biphoton quantum source generated in a lithium niobate microresonator system.
Without introducing the conventional domain poling, the on-chip microdisk produces entangled photon pairs covering a broad bandwidth promised by natural phase matching in spontaneous parametric down conversion.
Experimentally, the multiplexed photon pairs are characterized by $30\ \rm nm$ bandwidth limited by the filtering system, which can be furthered enlarged.
Meanwhile, the generation rate reaches $5.13\ {\rm MHz}/\upmu \rm W$ with a coincidence-to-accidental ratio up to $804$.
Besides, the quantum source manifests the prominent purity with heralded single photon correlation $g_H^{(2)}(0)=0.0098\pm0.0021$ and  energy-time entanglement with excellent interference visibility of $96.5\%\pm1.9\%$. 
}

\vspace{6pt}
\noindent
\textbf{Introduction}\\
\noindent
\noindent Integrated quantum photonics combining micron-scale fabrication technique and quantum information carriers promises a new platform for quantum computation and communication systems~\cite{wang2020integrated,feng2020progress,pan2012multiphoton}. 
For global quantum information processes, a systematic on-chip quantum device includes quantum source~\cite{guo2017parametric,steiner2021ultrabright}, quantum detector~\cite{sprengers2011waveguide,khasminskaya2016fully} and quantum transducer~\cite{mirhosseini2020superconducting}, \textit{etc}. 
Among them, quantum sources lay the foundation for all integrated quantum networks~\cite{bunandar2018metropolitan}, especially those capable of generating massive amount of qubits~\cite{zhong2020quantum}. 
So far, this target has been realized by raising the number of entangled photons or their dimensionality ~\cite{zhang2019generation,llewellyn2020chip,feng2016chip,wang2018multidimensional,feng2019generation}, technically fulfilled with an array of quantum sources~\cite{li2020metalens} or post selection of multiple photon pairs~\cite{luo2015quantum}. 
Nevertheless, entangled photon pairs with multiplexing ability provide an alternative to meet this requirement and overcome the possible incompatibility between different on-chip implementations~\cite{kues2017chip,reimer2016generation}.

Parametric processes including spontaneous parametric down conversion (SPDC) and spontaneous four wave mixing~\cite{lu2019chip,imany201850,feng2019chip} are normally used to generate entangled photon pairs~\cite{orieux2017semiconductor}. 
These approaches require materials possessing $\upchi^{(2)}/\upchi^{(3)}$ nonlinearity, for which lithium niobate (LN), aluminum gallium arsenide (AlGaAs), silicon nitride (SiN), silicon (Si), \textit{etc}. are appropriate candidates. 
Those $\upchi^{(3)}$ materials features compatibility with existing CMOS process while lack of conversion efficiency~\cite{llewellyn2020chip}.
And among $\upchi^{(2)}$ materials, LN is highly qualified for its superior nonlinear property~\cite{weis1985lithium,saravi2021lithium} as well as potential in integration.
With the commercialization of lithium niobate on insulator (LNOI) wafer and development of micro-fabrication technique, high-quality microstructures on LN chips are achieved aiming at promoting conversion efficiency and degree of integration~\cite{wang2014integrated,lin2015fabrication,zhang2017monolithic,lin2020advances}.
For the past two decades, versatile photon pair sources have been realized on LN systems such as bulk~\cite{wu1986generation}, waveguides~\cite{jin2014chip,xue2021ultrabright} and resonators~\cite{furst2010low,furst2011quantum,fortsch2013versatile} based on monocrystalline or periodically poled LN (PPLN)~\cite{yoshino2007generation,clausen2014source}.
Among diverse geometric designs, disk-shaped whispering gallery mode microresonators applied on LNOI remarkably enhance the light-matter interaction for the prominent high quality factors~\cite{vahala2003optical,lin2019broadband,lin2016phase,furst2016whispering,ye2020sum}.
Previous works devoted to LN microersonators either enhance the brightness of the quantum sources~\cite{ma2020ultra} or extend their bandwidth~\cite{luo2017chip}, leaving the ideal candidate for multiplexing quantum network still opening.

\begin{figure*}[ht]
    \centering
    \includegraphics[width=18cm,height=10.2cm]{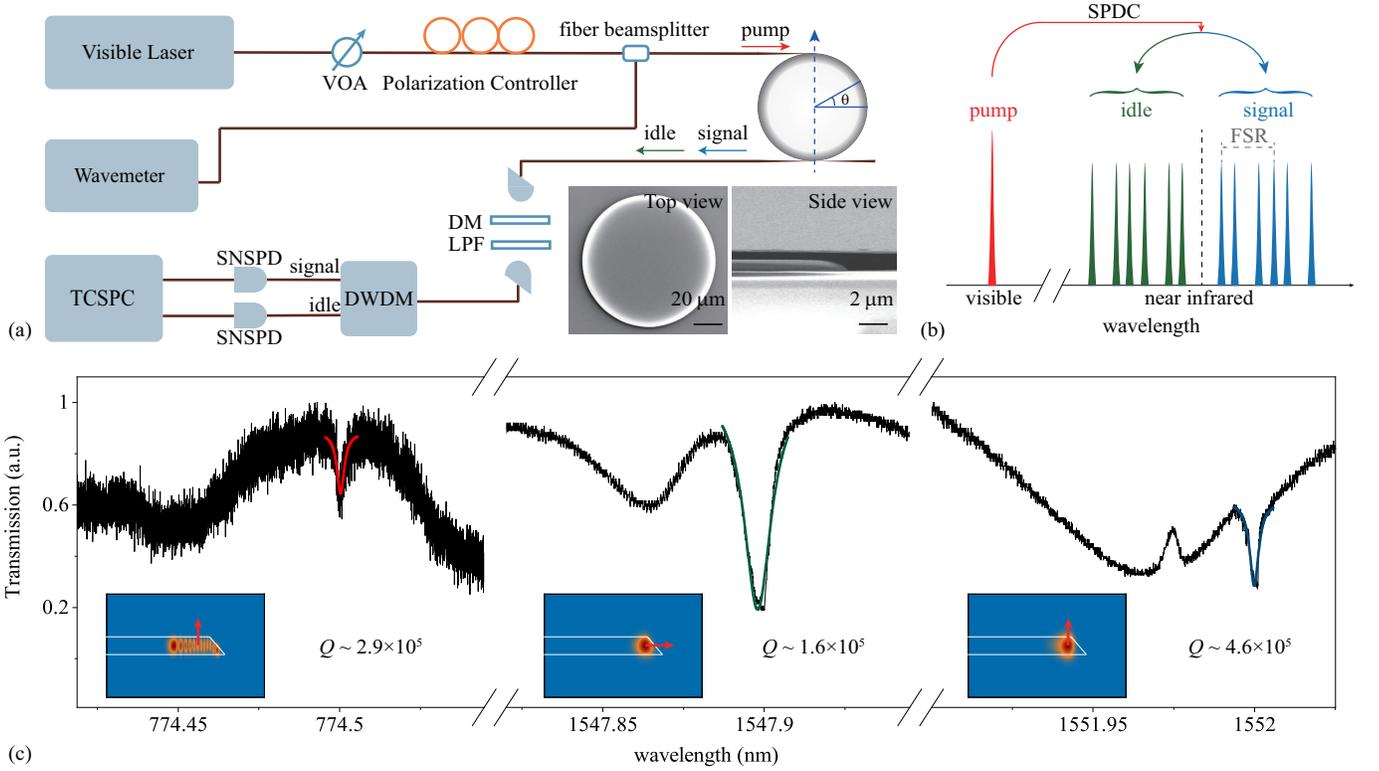}
    \caption{(a) Experimental setup for quantum detection. The blue arrow in the microresonator denotes the crystal axis and $\theta$ is the azimuthal angle. Insets: SEM images of LN microdisk. (b) Schematic of generating broadband entangled photon pairs via non-degenerate SPDC. (c) Mode charcaterization at visible and near-infrared band. Insets: field distributions of corresponding modes, red arrows denote the polarization. VOA: variable optical attenuator; DM: dichroic mirror; LPF: $1400\ \rm nm$ long pass filter; SNSPD: superconducting nanowire single photon detectors; TCSPC: time-correlated single photon counting.}
    \label{fig:my_label1}
\end{figure*}

Here, we experimentally demonstrate photon pairs generation on a high-quality LN microdisk not only having ultra-high brightness but also covering the full range of a commercial dense wavelength division multiplexing (DWDM) from $1535\ {\rm nm}$ to $1565\ {\rm nm}$. 
The birefringence of LN makes the propagation constant of transverse electric (TE) modes in the X-cut microdisks non-uniformed along the periphery, resulting in the azimuthally oscillation of phase mismatching around zero point. 
Consequently, the phase mismatching do not need to be equal to zero, which promises a natural broad phase-matching window. 
In our system, visible pump light at $775\ {\rm nm}$ and orthogonally polarized photon pairs around $1550\ {\rm nm}$ are coupled through two distinct tapered fibers for ideal extraction efficiency.
Such a concise but high-efficient platform free of PPLN structure ensures an superior pair generation rate (PGR) over $5.13\ {\rm MHz}/\upmu \rm W$ as well as unprecedented purity of source with a coincidence-to-accidental ratio (CAR) up to $804$ and heralded single photon correlation (HSPC) $g_H^{(2)}(0)=0.0098\pm0.0021$.
Moreover, the photon pair source is verified for energy-time entanglement with an excellent interference visibility of $96.5\%\pm1.9\%$.
This versatile photon pair source possessing both high PGR and wide bandwidth paves the way for a broad spectrum quantum information process on chip.

\vspace{6pt}
\noindent
\textbf{Results}\\
\noindent \textbf{Experimental configuration.} The experimental setup has been illustrated in Fig.~\ref{fig:my_label1}(a). 
In preparation, the microdisk was fabricated on LNOI by photolithography-assisted chemo-mechanical etching~\cite{wu2018lithium} and the details are listed in the device fabrication part of the Method.
A freestanding LN microdisk with a diameter of 93 $\upmu \rm m$ possessing a smooth sidewall and a wedge angle of approximately $35\degree$ is finally applied, as shown in the scanning electron microscopy (SEM) images of Fig.~\ref{fig:my_label1}(a). 
In experiment, the visible pump light from a continuous wave laser is launched into the LN microdisk via a tapered fiber. 
Meanwhile, the generated signal and idle lights are collected from another tapered fiber specially designed for coupling at the communication band with clockwise and counter-clockwise collection ratio around $20:1$, which is a great promotion of previous work~\cite{luo2017chip}.
The coupling efficiencies of the pump and collection fiber are optimized prior to the collection rate of signal and idle lights, resulting in $0.363\%$ at $774.86\ \rm nm$ pumping mode and up to $55.65\%$ for $1550\ \rm nm$ fundamental modes respectively.
The averaged coupling efficiency for the modes at $1550\ \rm nm$ band is estimated as $25\%$, which is pulled down by the poorly coupling of the modes with high radial mode numbers.
In case of phase matching condition and energy conversation being fulfilled, the non-degenerate SPDC process will happen and a single pump photon will be converted into a signal photon and an idle photon.
As illustrated from Fig.~\ref{fig:my_label1}(b), the broadband photon pairs are composed of two parts: one are the modes from different mode families within a free spectral range (FSR) and the other are the modes from the same mode family with an FSR interval.
Practically, as shown in Fig.~\ref{fig:my_label1}(c), the non-degenerate SPDC photon pairs generate from a transverse magnetic (TM) mode at $774.86\ \rm nm$ with high radial modal number to two corresponding orthogonally polarized modes with low radial modal numbers.
Acquired from the transmission spectra in Fig.~\ref{fig:my_label1}(c), the FSR is estimated as $3.89\ \rm nm$ and $3.67\ \rm nm$ for signal and idle light respectively for the applied microdisk.
Meanwhile, the loaded quality ($Q$) factor is $2.9\times 10^5$ at pump light wavelength and universally around $10^5$ at signal and idle light wavelengths.
It should be noted that the severe loss of $Q$ factors from intrinsic $Q$ over $10^7$ blames on the great scattering decline from the adhesion of fiber and microdisk for the sake of system stability. 
Explicitly, details on the characterization on the coupling efficiencies and transmission spectra are shown in the optical measurement part of the Method.

\begin{figure}[ht!]
    \centering
    \includegraphics{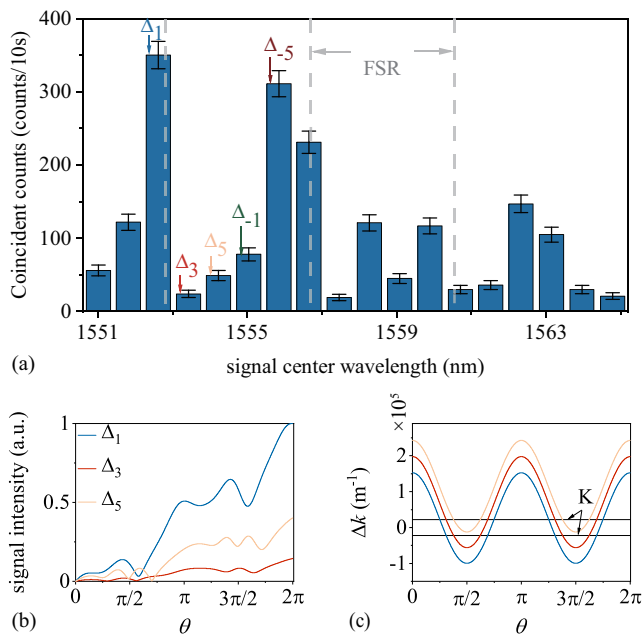}
    \caption{(a) Spectrum of coincident counts presented by the signal light side at different channels of a DWDM. The numbers denote different mode families with different phase-matching azimuthal modal number conditions $\Delta m$. (b) The growth of signal light intensity from different mode families along the azimuthal angle $\theta$ of the microresonator. (c) The oscillation of phase mismatching quantity $\Delta k$ along $\theta$.}
    \label{fig:my_label2}
\end{figure}

The signal is filtered by a dichroic mirror and a successive long pass filter before entering the superconducting nanowire single photon detectors (SNSPD), ensuring the elimination of pump light.
Especially, the coincident counts spectrum is gathered through a DWDM with $0.8\ \rm nm$ resolution at $46.5\ \upmu\rm W$ pump power shown in Fig.~\ref{fig:my_label1}(a).
The spectrum is demonstrated at the signal light side of the photon pairs, and the maximal coincident counts reach $350$ in a $10\ \rm s$ time integral with center wavelength at $1552.52\ \rm nm$ shown in Fig.~\ref{fig:my_label2}(a).
The loss on the signal and idle light is estimated as $25\ \rm dB$ from the generation to detection.
Precisely, the losses are calibrated as $2.2\ \rm dB$, $2.4\ \rm dB$ and $4\ \rm dB$ on the filtering opponents including the LPF, DM and DWDM.
Moreover, the losses on the connection between different devices mainly the adapters and the transmission in the tapered fiber are $8.7\ \rm dB$ and $3\ \rm dB$, respectively.
It should be noted that the shrink of coincident counts at longer wavelength results from the existence of material dispersion. 
In the longer wavelength, the deviation in the frequency between the signal lights and the resonant modes grows to exceeding the linewidth of the modes.  
Limited by the range of the DWDM, the spectrum of photon pairs is verified within $1535\ {\rm nm}$ to $1565\ {\rm nm}$, which is supposed to be broader than $200\ \rm{nm}$ with the appearance of signal at the border of the DWDM~\cite{luo2017chip}. 

\vspace{6pt}
\noindent \textbf{Phase matching technique.} Such a broadband photon pair spectrum is generated through natural phase-matching condition with the assistance of the birefringence and the azimuthal modulation on nonlinear effective index. 
The parametric process can be described in the quantumn representation to analyze the signal intensity as follows, and the derivation is listed in the Method.
\begin{equation}
\begin{split}
P=\sum_{k_s,q}\sum_{k_i,q}A\iiint \hat a_p(\omega_p)\hat  a_{s,q}^\dagger(\omega_s)\hat a_{i,q}^\dagger(\omega_i)\\
\chi ^{(2)}(\theta) e^{i(k_s+k_i-k_p)Rd\theta}rdrd\theta dz .   
\end{split}
\label{my_equ5}    
\end{equation}
Specifically, the second order nonlinear coefficient $\chi^{(2)}$ can be subjected to effective nonlinear coefficient with a fixed choice of involved light polarization. 
In our system, it goes with $d_{\rm eff}=d_{22}\cos{\theta}+d_{31}\sin{\theta}$ for two TM-polarized lights and one TE-polarized light involvement.
To elucidate the experiment result, we employ a simulation model that has the same geometrical scale with the realistic microdisk for finite element method calculation.
Naturally, the refractive index of TE-polarized light experiences an oscillation along the azimuthal angle due to the birefringence with $n_{\rm TE}=1/\sqrt{\cos^2{\theta}/n_o^2+\sin^2{\theta}/n_e^2}$, where $n_o$ and $n_e$ are the ordinary and extraordinary refractive index of LN~\cite{lin2019broadband}.
Thus, the phase mismatching quantity $\Delta k$ oscillates in the same way. 
Meanwhile, the spatially modulated effective nonlinear coefficient $d_{\rm eff}$ can be written in the form of Fourier expansion with first order series, which leads to a phase compensation term $K$. 
Considering the resonant condition of microdisk, the wave vector satisfies $n_{\rm eff}kR=m$ where $m$ is the azimuthal modal number. 
Namely, the phase mismatching quantity can be denoted as $\Delta m$, and characterizes the phase-matching conditions of different mode families shown in Fig.~\ref{fig:my_label2}(a).
The maximal coincident counts originated from two fundamental modes at photon pair band illustrated in Fig.~\ref{fig:my_label1}(c) with $\Delta m=1$.
The growth of signal intensity in single circle integral of $P$ is shown in Fig.~\ref{fig:my_label2}(b), and the intensity varies with different cases of $\Delta m$.
It is obvious that the electric field intensity overlap integral as well as phase difference accumulation make joint contributions to the output signal intensity.
Besides, the overlap between $\Delta k$ and $K$ promises persistent accumulation of signal intensity beyond one circle.
Such requirement also restricts the choice of $\Delta m$ in the range of $\left[-5,5\right]$ partially shown in Fig.~\ref{fig:my_label2}(c).
It should be noted that the resolution window of DWDM empirically contains multiple modes of the microdisk, thus the identification of mode families in Fig.~\ref{fig:my_label2}(a) is a bit rough.
The phase matching window can be estimated by the comparison of the modal frequency mismatching $\Delta f=f_s+f_i-f_p$ and the modal linewidth (approaximately $\rm{0.3\ GHz}$).
As a simulation result, $\Delta f$ is within the linewidth with the appearance of dispersion for the signal light eigenmode shown in Fig.~\ref{fig:my_label1}(c) at $\rm{1552-1647\ nm}$ band, resulting in a nearly $\rm{200\ nm}$ photon pair generation bandwith in theory. 
Nevertheless, the dynamical tuning from the fiber-resonator coupling and the nonlinear process like thermal and photorefractive effect deviates the modes from the calculated eigenmodes, which narrow down the phase matching window in experiment.

\begin{figure}[t!]
    \centering
    \includegraphics{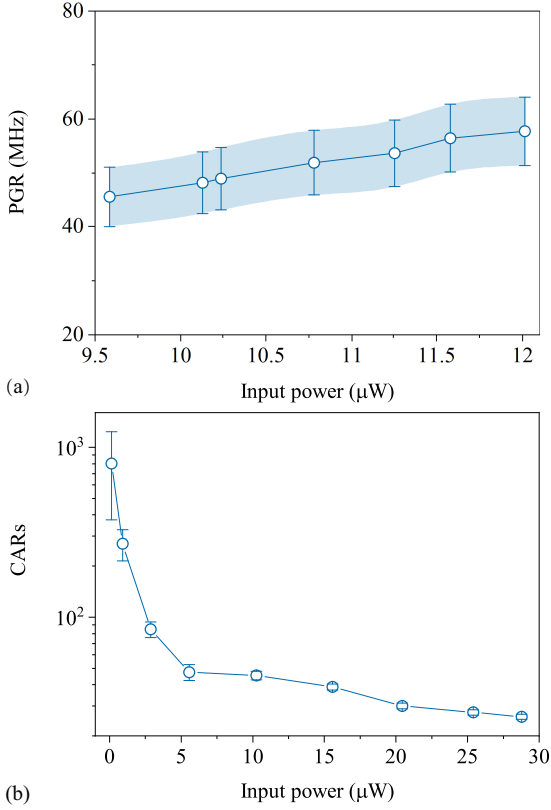}
    \caption{(a-b) Power dependence of PGR and CAR. The coincident counts are measured under the $85\%$ efficiency of SNSPD and dark count rate below $100\ \rm {Hz}$. The error bar is calculated by $\sqrt{N}.$}
    \label{fig:my_label3}
\end{figure}

\vspace{6pt}
\noindent \textbf{Quantum characterization.} To characterize the quantum source, we record the collective PGR and CAR free of DWDM filtering and the results are shown in Fig.~\ref{fig:my_label3}(a-b).
Considering the lifetime of photons in the applied resonator around $200\ \rm ps$, we set the coincident window to be $0.8\ \rm ns$ to cover the coincident event.
Typically, the PGR is calculated as ${\rm{PGR}}={N_1N_2}/{N_{12}}$, where $N_1$, $N_2$ and $N_{12}$ are the counts for single detector and coincidence respectively.
The efficiency of SPDC is fitted as $5.13\ {\rm MHz}/\upmu \rm W$ and the maximal output of photon pairs flux reaches $136.5\ \rm MHz$ with the pump power $27.3\ \upmu\rm W$. 
In practice, the raise of intracavity power renders dramatical resonant mode detuning especially at visible band due to the thermal and photorefractive effect~\cite{wang2016thermo,sun2017nonlinear}, resulting in a limitation on PGR.
The CAR reaches up to $804$ at a low pump power $0.14\ \upmu\rm W$ and decreases as the pump power increases, which mainly attributes to the increase in multi-pair generation.

\begin{figure}[ht]
    \centering
    \includegraphics{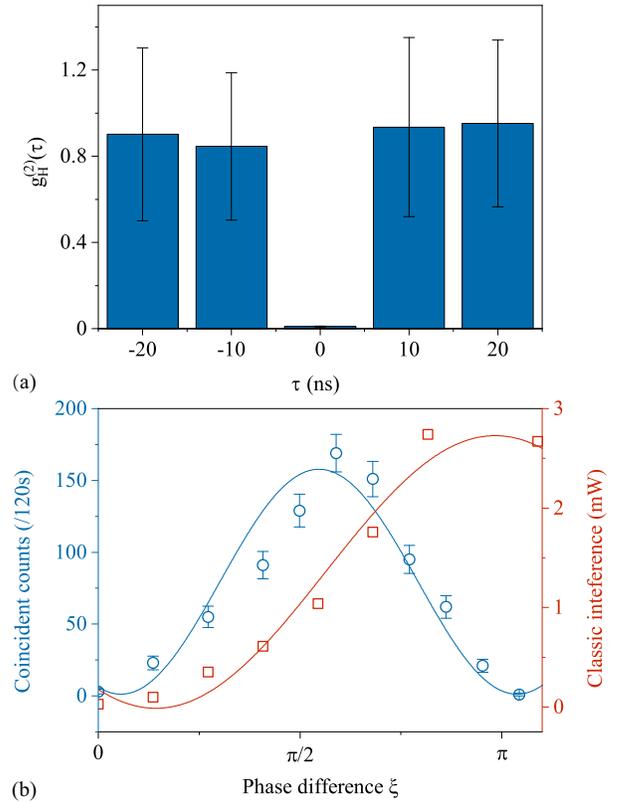}
    \caption{(a) Single photon purity verification for HSPC function $g^{(2)}_{H}(\tau)$. Integral time: $1\ \rm h$. (b) Two-photon energy-time entanglement measurement. Comparison between quantum and classical interference behaviors. Integral time: $120\ \rm s$.}
    \label{fig:my_label4}
\end{figure}

Furthermore, we testify the single photon purity by measuring the HSPC function under a Hanbury Brown and Twiss experiment.
The measurement is conducted under the condition that the corresponding DWDM channels possess the highest coincident counts with collective $\rm{PGR}=70.2\ \rm MHz$.
The correlation function is gathered by $g^{(2)}_{H}(\tau)={N_{is_1s_2}N_i}/{N_{is_1}N_{is_2}}$, where $N_i$ is the counts of idle photons, $N_{is_1}$, $N_{is_2}$ and $N_{is_1s_2}$ are the coincident counts between two-path and three-path respectively. 
$\tau$ is the delay time of the electronic trigger pulses from one detector for signal photons with respect to the idle photons.
As illustrated in Fig.~\ref{fig:my_label4}(a), the shape of $g_H^{(2)}$ forms an anti-bunching dip with an impressively low $g_H^{(2)}(0)=0.0098\pm 0.0021$, indicating an excellent single photon quality as an ultrabright quantum source.

Besides, the energy-time entanglement analysis is exhibited with the aid of a common unbalanced Michelson interferometer (UMI) with $1.6\ \rm {ns}$ time difference. 
Two-photon state is modulated in time, and three coincidence peaks are displayed in the TCSPC system with a time window of $0.8\ \rm{ns}$. 
By post-selection of the central peak of the coincident counts with a proper time window, the state after post-selection is $|\Phi\rangle=\frac{1}{\sqrt{2}}(|SS\rangle+e^{2i\xi}|LL\rangle)$, where the first and second terms denote both signal and idle photons passing the short (S) and long (L) arms of the UMI, and $\xi$ is the phase difference introduced by UMI. 
Two-photon interference fringe is given in Fig.~\ref{fig:my_label4}(b), where a comparison between classical and quantum behaviors is also exhibited. 
Compared with classic coherent light from laser, it is clear that the interference period of classical light doubles that of quantum entangled light.
Finally, the interference visibility is estimated as $96.5\%\pm 1.9\%$ for the entangled state.

\vspace{6pt}
\noindent
\textbf{Discussion}\\
\noindent
Disk-shaped microresonators naturally support abundant mode families, which is an appropriate platform to accomplish broadband source generation through natural phase-matching condition.
Besides, the considerable quality factor of this microdisk ensures a concise and robust system to fulfill resonant condition with broad bandwidth as well as high efficiency.
As a result, a remarkable PGR at $5.13\ {\rm MHz}/\upmu \rm W$ and a spectral bandwidth over $30\ \rm nm$ limited by the DWDM are experimentally demonstrated. 
That is an excellent candidate for multi-channel frequency-encoded quantum computation and quantum communication process.
In Table.~\ref{Table.1}, we compare the performance of reported devices with different materials including $\rm SiN$, LN, Si, AlN and AlGaAs. 
To our knowledge, the experimental configuration we present is unique in achieving ultra-high brightness and considerably broad bandwidth simultaneously, and the bandwidth can be further broadened towards the theoretical value $\rm{200\ nm}$ with the promotion of the filtering devices.
Meanwhile, lack of phase mismatching between different materials, the coupling efficiency of fiber-resonator system for cross-band coupling can be further optimized. 
Future designs including chaos-assisted broadband coupling~\cite{jiang2017chaos} and integrated waveguide coupling may resolve this obstacle.
Moreover, a multi-photon source can be promisingly acquired by switching the pump laser from a continuous wave to a pulsed one~\cite{zhang2019generation}. 
In conclusion, combination of the multiplexing application and great brightness, together with the breakthrough of micron-scale footprint make our entangled photon source set a bright future for integration of quantum networks.

\vspace{6pt}
\noindent
\textbf{\large Methods}\\
\noindent
\textbf{Device fabrication.} The microresonator used in our experiment is fabricated on a comercially available X-cut LN thin film wafer (NANOLN, Jinan Jingzheng Electronics Co., Ltd.)  with a thickness of $900\ \rm nm$ bonded to a $2\ \rm \upmu m$ thick $\rm{SiO_2}$ layer grown on a LN substrate. 
The fabrication process can be summarized in four steps. 
First, the magnetic sputtering is applied to deposit a layer of chromium (Cr) with a thickness of $600\ \rm nm$ on the surface of the LNOI. 
Second, through the space-selective femtosecond laser direct writing, the Cr layer is patterned into a circular disk with high precision which will be used as a hard mask on the LNOI for the next polishing procedure. 
In this step, a femtosecond laser centered at $\rm{1030\ nm}$ with the repetition rate of 250 kHz and pulse width of 190 fs is utilized. 
Thanks to cautiously pulse energy choice and low height generation during fabrication, the Cr mask is shaped without damaging the underneath LN film. 
Third, the chemo-mechanical polishing (CMP) is carried out to transfer the pattern of the Cr hard mask into the thin film LN via etching the exposed thin film. 
In the last step, the Cr mask is removed and the $\rm{SiO_2}$ layer is undercut into a small pedestal for supporting the LN microdisk by a two-step chemical etching with buffered hydrofluoric acid solution. 
As a result, an X-cut LN microdisk supported by the $\rm{SiO_2}$ pedestal with a diameter of $93\ \rm{\upmu m}$ is achieved, whose sidewall presents sub-nanometer smooth with a wedge angle of approximately $35^\circ$.

\vspace{6pt}
\noindent
\textbf{Optical measurement.} In the optical characterization of our device, the LN microdisk is coupled to a tapered fiber with the ability of precisely controlling the coupling condition. 
The transmission spectra at the $1550\ \rm nm$ and $780\ \rm nm$ are respectively acquired by pumping wavelength scanned lasers into the LN microdisk (Santec TSL-510 for $1500- 1630\ \rm nm$ and New Focus TLB-6700 for $765-780\ \rm nm$) through the tapered fiber. 
In the experiment, the polarization of pumping light is controlled by a three-paddle polarization controller to achieve the required states. 
To verify the polarization state, the emission light of the microdisk is tested with a linear polarizer. 
In the experimental setup, a visible fiber tip and an infrared tapered fiber are simultaneously applied as the pumping and collection fiber.
The pumping efficiency is estimated by monitoring the power of the coupled visible light in the pumping fiber when pumped in the collection fiber, and the averaged collection efficiency is estimated by the optical losses.
Moreover, the clockwise and counter-clockwise collection ratio is estimated by the comparison of coincidence counts at each direction, in which the counter-clockwise part results from the backscattering in the microdisk.
Finally, a visible light from a continuous wave laser (Toptica 780 DL pro) is used to pump the quantum source, whose wavelength is monitored by a wavemeter with a fiber beamsplitter before the coupling.  

In the quantum characterization parts, the PGR and CAR is measured for photons of full-bandwidth when the entanglement and correlation for photons within single DWDM channel. 
In the entanglement measurement, the UMI is modulated by changing the temperature of one arm of the interferometer.  
First of all, we collected the interference curve of a classic laser with wavelength of $1550\ \rm nm$ to determine the adjustment range of temperature within which the interference consequence was stable and the classic result as well. 
Then, the UMI was inserted between the DWDM and spacial filter system for the two-photon energy-time entanglement measurement. 
The three peaks in the TCSPC system  separately represent two photons passing both short arms, one arm each and both long arms, among which the middle peak is needed.

\vspace{6pt}
\noindent
\textbf{Parametric process in quantum representation.} To exhibit the mechanism of broadband SPDC process in our regime, the interaction Hamiltonian of SPDC involved with pump light $\hat E_p$, signal light $\hat E_s$ and idle light $\hat E_i$ is introduced as~\cite{yang2008spontaneous,gong2011compact}
\begin{equation}
H_I(t)=\epsilon_0 \int_V \chi ^{(2)}(\vec {r},t)\hat E_p^{(+)}(\vec {r},t)\hat E_s^{(-)}(\vec {r},t)\hat E_i^{(-)}(\vec {r},t)d^3\vec{r}+H.c.
\label{my_equ1}    
\end{equation}
The continuous TM-polarized pump laser can be treated as plane-wave in the form of
\begin{equation}
\hat E_p^{(+)}(\vec {r},t)=E_pe^{-i\vec{k}_p\cdot\vec{r}+i\omega_pt}\hat a_p(\omega_p).
\label{my_equ2}    
\end{equation}
And the orthogonally polarized versatile signal and idle lights can be treated as ($q$ denotes the polarization)
\begin{align}
& \hat E_s^{(-)}(\vec {r},t)=\sum_{\vec{k}_s,q}E_{s,q}^*e^{i\vec{k}_s\cdot\vec{r}-i\omega_st}\hat a_{s,q}^\dagger(\omega_s),\\
& \hat E_i^{(-)}(\vec {r},t)=\sum_{\vec{k}_i,q}E_{i,q}^*e^{i\vec{k}_i\cdot\vec{r}-i\omega_it}\hat a_{i,q}^\dagger(\omega_i),
\label{my_equ3}
\end{align}
where $E_p$, $E_{s,q}$ and $E_{i,q}$ are the quantized coefficients represented by $E_j=i\sqrt{\hbar \omega_j/(4\pi\epsilon_0 c n_j)}$, $j=p,s,i$. $\hat a_p(\omega_p)$, $\hat a_{s,q}^\dagger(\omega_s)$ and $\hat a_{i,q}^\dagger(\omega_i)$ are the annihilation and creation operators of pump, signal and idle lights, respectively. In the weak signal approximation of SPDC process, the status of system can be represented by
\begin{equation}
\left |\Psi \right \rangle=\left |vac \right \rangle+\frac{1}{i\hbar}\int_{-\infty}^{\infty}H_I (\tau)\left |vac \right \rangle d \tau .
\label{my_equ4}
\end{equation}
Then, by substituting Eq.~\ref{my_equ1} to Eq.~\ref{my_equ4}, the generation intensity of signal as Eq.~\ref{my_equ5} can be obtained in the cylindrical coordinates under the conservation of energy.



\vspace{6pt}
\noindent \textbf{Acknowledgment}
\noindent
We thank W. Liu for helpful discussion. This project is supported by the National Key R$\&$D Program of China (Grant No. 2016YFA0301302 and No. 2016YFA0301700) and NSFC (Grants Nos. 11825402, 61590932, 11774333, 62061160487, 12004373, 11734009, 11874375), the Anhui Initiative in Quantum Information Technologies (No. AHY130300), the Strategic Priority Research Program of the Chinese Academy of Sciences (No. XDB24030601), the Beijing Academy of Quantum
Information Sciences (Grant No. Y18G20) and the Fundamental Research Funds for the Central Universities. This work was partially carried out at the USTC Center for Micro and Nanoscale Research and Fabrication.

\vspace{6pt}
\noindent \textbf{\large Author contributions}
\noindent
\\X.-F.R., Y.-F.X. and Y.C. conceived the idea and designed the experiments. C.-H.D and H.-Y.Z provided guidance and experimental instruments for this work. B.X. performed experiments, measurements and data collection with assistance of L.F. and R.N.. J.L. fabricated the device with the assistance of R.G.. L.-K.C. carried out the theoretical analysis and wrote the manuscript with contribution from all authors. X.-F.R., Y.-F.X., Y.C., G.-C.G. and Q.G. supervised the project.

\vspace{6pt}
\noindent \textbf{\large Competing interests}
\\The authors declare no competing interests.


\end{document}